\documentclass[prl,twocolumn,showpacs,showkeys]{revtex4}
\usepackage{amsmath,graphicx,amsfonts,bm,amssymb}
\usepackage{times}

\begin{document}

\title{Producing GHZ state of nitrogen-vacancy centers in cavity QED}

\author{Zheng-Yuan Xue}

\author{Sheng Liu}

\affiliation{Laboratory of Quantum Information Technology, and School of Physics and
Telecommunication Engineering, South China Normal University, Guangzhou 510006, China}

\date{\today}

\begin{abstract}
We propose a  scheme to  generate GHZ state of nitrogen-vacancy centers coupled to a
whispering-gallery mode cavity. The implemented evolution is independent on the cavity field
state with the assistant of a strong classical field, and thus not sensitive to the thermal
state of the cavity. Meanwhile, it is fast compare to the convectional dispersive interaction
induced operation in a cavity-assisted system. The scheme is readily scalable to multiqubit
scenario.
\end{abstract}

\pacs{03.67.Lx, 42.50.Dv}

\keywords{cavity QED; GHZ state; Nitrogen-vacancy centers}

\maketitle

Entanglement is one of the most counterintuitive properties of quantum mechanics, which plays
an central role in quantum information and quantum computation \cite{e}. For real quantum
system, it is generally sensitive to practical noise, which results from the infamous
decoherence effect. Therefore, generation of entangled state with inherent robustness to
diverse noise is preferred. Due to  long electronic spin lifetime, fast initialization, good
qubit readout, and coherent manipulation at room temperature, the  diamond nitrogen-vacancy
(NV) center is considered as a promising candidates for quantum computing \cite{To}. Coupling
of many qubits in solid state systems is problematic. For NV center, the partial solution is
introduce coherent coupling between electron and nuclear spin qubits. However, this can only be
done in the case of a few qubits since there are limited number of nuclear spins can be
individually addressed \cite{N}. Alternatively, separated NV centers can be coupled by a cavity
mode \cite{cavity}. For a easy fabricated microsphere cavity, the quantized whispering-gallery
mode (WGM) may has a ultrahigh Q factor and very small volume \cite{q}, and thus leading to
reach the strong coupling regime. Entanglement generation schemes are also proposed  with a
microsphere cavity \cite{generation} or a one dimensional resonator \cite{1d}. However, the
previous schemes all work in the dispersive regime, i.e., the detuning between the cavity mode
frequency and the qubits is much larger than that of the cavity-qubit interaction, which makes
the effective coupling relatively  weaker compare to resonant coupling, that is, the generation
time will be longer. Meanwhile, the coupling usually contains cavity-state-dependent energy
shift, which leads the generation to be sensitive to the cavity thermal state.

Here, we propose an alternative scheme for generating multipartite entangled state of
nitrogen-vacancy centers coupled to a whispering-gallery mode cavity, which overcome the
above-mentioned drawbacks. In addition to the optical transitions driven by the cavity mode and
optical laser, we introduce another transition within the ground states strongly driven by a
microwave field. Before proceeding, we want to emphasize that using both optical and microwave
radiation to control an electron spin associated with the NV center in diamond is
experimentally accessible \cite{om}. With the  driven of a microwave field, one get two
additional merits of the proposed scheme. Firstly, the photon-number-dependent parts in the
evolution operator are canceled, and thus the scheme is insensitive to the thermal state of the
cavity mode. Secondly, the large detuning constrain can be removed so that fast gate operation
can be achieved. Meanwhile, the solid-state set-up is readily scalable to multiqubit scenario
and the gate speed is not slowed down with the increasing of the involved qubits.

As illustrated in Fig. 1(a), $N$ identical NV centers in $N$ separate diamond nanocrystals may
be strongly coupled to the WGM of a microsphere cavity.  The lowest-order WGM corresponding to
the light traveling around the equator of the microsphere, which offers exceptional mode
properties for reaching strong coupling regime. One can model the NV center as a three-level
system, as shown in Fig. 1(b), where the states $\left\vert ^{3}A,m_{s}=0\right\rangle $ and
$\left\vert ^{3}A,m_{s}=-1\right\rangle $ are encoded as our qubit states $\left\vert
0\right\rangle $ and $\left\vert 1\right\rangle $, respectively. The state $\left\vert
^{3}E,m_{s}=0\right\rangle $ is labeled by $\left\vert e\right\rangle $ and do not use the
metastable $^{1}A$ state, which has not yet been fully understood \cite{Rog}. In our
implementation, the optical transition $\left\vert 0\right\rangle \rightarrow \left\vert
e\right\rangle$ and $\left\vert 1\right\rangle \rightarrow\left\vert e\right\rangle $ (with
transition frequencies $\omega_{e0}$ and $\omega_{e1}$ ) are coupled by the WGM with frequency
$\omega _{c}$ and a laser with frequency $\omega _{L}$ and polarization $\sigma ^{+}$
\cite{lambda}, respectively. Both coupling is far-off resonant from their transition
frequencies  so  that the $|e\rangle$ state can be adiabatically eliminated. The NV centers are
fixed and apart with the distance much larger than the wavelength of the WGM, so that they can
interact individually with the laser beams and the direct coupling among them is negligible.
Then, in units of $\hbar =1$, the system is described by
\begin{eqnarray}
H_{S}&=&\omega_c a^{+} a +  \sum_i \omega_i |i\rangle\langle i| \\
&&+\sum _{j=1}^{N}[G _{j} a|e\rangle\langle 0|+
\Omega_{L,j}e^{-i(\omega_L+\phi)}|e\rangle\langle 1| + \text{H.c.}],\notag
\end{eqnarray}
where $a^{+}(a)$ is the creation (annihilation) operator of the WGM field, $\omega_i$ is the
frequency of $i$th energy level with $i\in\{0, 1, e\}$, $G_j$ and $\Omega_{L,j}$ is the
coupling strength between $j$th NV center and WGM and laser, respectively.

\begin{figure}[tbp]
\centering
\includegraphics[width= 8cm]{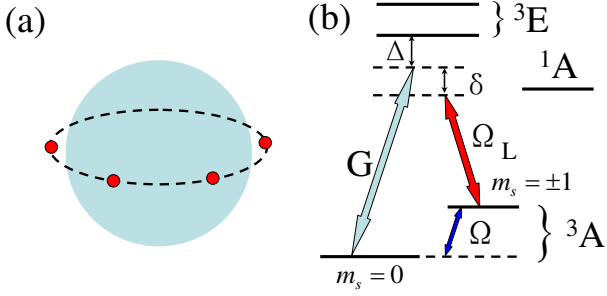}
\caption{(a) Schematic setup of the microsphere cavity, where
$N$ identical NV centers (red dots) in diamond nanocrystals are
equidistantly attached around the equator of a single fused-silica
microsphere cavity. (b) Level diagram for a NV center, with $G$ and
$\Omega _{L}$are the coupling strength between NV center and WGM and
the laser pulse, respectively. We encode qubits in the subspace
spanned by the state of $m_{s}=0$ and $m_{s}=-1$. The qubit
zero-field splitting is $\omega_{10}=$2.88 GHz and can be driven by
a microwave field with strength $\Omega$.  }
\end{figure}

Using the rotating-wave approximation (RWA), with respect to the free Hamiltonian $H_0=\omega_c
a^{+} a +  \sum_i \omega_i |i\rangle\langle i| $, the interaction Hamiltonian in the
interaction picture can be written  as \cite{eh}
\begin{equation}
H_{I}=\sum _{j=1}^{N}\eta _{j}\left(a\sigma _{j}^{+}e^{-i (\delta_{j} t -\phi)}+\text{H.c}.\right),
\end{equation}
where   $\sigma _{j}^{+}=\left\vert 1_{j}\right\rangle \left\langle
0_{j}\right\vert $, $\sigma _{j}^{-}=\left\vert 0_{j}\right\rangle
\left\langle 1_{j}\right\vert $, and $\eta _{j}=G_{j}\Omega _{L,j}(\frac{1%
}{\Delta _{j}+\delta _{j}}+\frac{1}{\Delta _{j}})$ with $\Delta _{j}=\omega _{e0,j}-\omega
_{c}$ and $\delta _{j}=\omega _{c}-\omega _{10,j}-\omega_{L,j}$.  For simplicity, hereafter, we
assume that   $\eta_{j}=\eta$ and $\delta _{j}=\delta.$ However, this is not necessary here.
The    inhomogeneous  qubit-cavity coupling will result in deviation of the generated final
state from the target state by   extra phase factor, which can be easily compensated by single
qubit operations \cite{dl}. Meanwhile, besides the optical coupling, we also add a resonant
microwave driven field at a frequency of the microwave transition $\omega_{10}$ for each NV
center. Then, in the above interaction picture, it reduce to $H_d={\Omega \over 2} \sigma_j^x$
with the Pauli matrix written in the qubit subspace. Then the total interaction Hamiltonian can
be written as
\begin{equation}
H_{1}=\sum _{j=1}^{N} \left[{\Omega \over 2} \sigma_j^x +
\eta\left(a\sigma _{j}^{+}e^{-i(\delta  t -\phi)}+a^{+}\sigma _{j}^{-}e^{i(\delta  t-\phi)}\right)\right].
\end{equation}
This microwave control over the NV centers is well studied experimentally \cite{omega} and even
all the three magnetic sub-levels can be resolved \cite{du}. With this strong microwave
coupling, we can realize the analogy of strong-driven atomic cavity QED system with two level
atoms \cite{qed}. Meanwhile, this distinct our scheme with all the previous investigation with
NV centers coupled by a cavity mode \cite{cavity,generation,1d}. In the interaction picture
with respect to the first term, the interacting Hamiltonian reads \cite{qed}
\begin{eqnarray}\label{Hint}
H_{2}&=& {\eta \over 2} a  e^{-i(\delta  t-\phi)} \sum_{j=1}^N
\left(\widetilde{\sigma}_{z}^j +e^{i\Omega
t}|+\rangle_j\langle-|-e^{-i\Omega
t}|-\rangle_j\langle+|\right)\notag\\
&& +\text{H.c}.
\end{eqnarray}
where $\widetilde{\sigma}_z$ denotes Pauli matrix in the eigen basis of $\sigma_x$ with eigen
equations as $\sigma_x|\pm\rangle=\pm|\pm\rangle$, where
$|\pm\rangle=(|0\rangle\pm|1\rangle)/\sqrt{2}$. In the case of $\Omega\gg\{\delta, \eta \}$, we
can omitting the fast oscillation terms with frequencies $\Omega\pm\delta$ using RWA, then the
Hamiltonian (\ref{Hint}) reduces to
\begin{eqnarray}\label{Hint1}
 H_{3}=  \eta  \left(a e^{-i(\delta  t-\phi)} + a^{\dag} e^{i(\delta  t-\phi)} \right) J_x,
\end{eqnarray}
where $J_x=\sum_{j=1}^N \sigma_j^x/2$. For $\phi=0$, the above Hamiltonian reduces to
\begin{eqnarray}\label{H0}
 H_{D1}=  \eta   \left(a e^{-i \delta  t } + a^{\dag} e^{i\delta t} \right) J_x.
\end{eqnarray}
The time-evolution operator for the Hamiltonian in Eq. (\ref{H0}) can be expressed as
\cite{ham,zhu,xue,ms}
\begin{eqnarray}\label{U0}
U(t) &=& \exp\left[{-iA(t)J _{x}^{2}}\right]\notag\\
&& \times \exp\left[{-iB(t)aJ_{x}}\right]
\exp\left[{-i{B^{\ast}}(t)a^{\dagger}J_{x}}\right],
\end{eqnarray}
where
\begin{eqnarray}
A(t)=\frac{\eta^{2}}{\delta}\left[  \frac{1}{i\delta}\left(
e^{i\delta t}-1\right)-t\right],
\end{eqnarray}
\begin{eqnarray}\label{b}
B(t)=i\frac{\eta}{\delta}\left(e^{-i\delta t} -1 \right).
\end{eqnarray}
While the whole time evolution is nonperiodic,  Eq. (\ref{b}) shows that $B(t)$ is a periodic
function of time, and it vanishes at $\delta T=2k\pi$ where $ k=1,2,3,...$. At these times, the
time evolution operator in the interaction picture reduces to
\begin{eqnarray}\label{u1}
U(\gamma')=\exp(i\gamma' J_{x}^{2}),
\end{eqnarray}
with $\gamma'=\eta^{2}T/\delta \equiv\lambda T$.

The operator in Eq. (\ref{u1}) is readily to be used for generating GHZ state. However, this
requires a carefully chosen duration of the interaction because it involves real excitation of
the cavity photon. Fortunately, idea similar to the spin echo  and  dynamical decoupling can be
used to optimize the generation process \cite{wang,monroe,zou}. The key is the map of
$H_{D1}\rightarrow H_{D2}=-H_{D1}$, which can be conveniently achieved by setting $\phi=\pi$ in
our scheme. Then, a gate
\begin{eqnarray}\label{ut}
U(\gamma)=\exp(i\gamma J_{x}^{2}),
\end{eqnarray}
with $\gamma=\frac{\eta^{2}}{\delta}\left({2\over \delta} \sin{\delta t \over 2}- t\right)$ can
be obtained  as the following  \cite{wang,monroe,zou}. For a gate with duration $t$, one can
set $H_{D1}$ for the duration of $0\sim t/2$ and $H_{D2}$ for the left duration $t/2 \sim t$
\cite{wang}, similar to the dynamic decoupling method. Alternatively, the map of
$H_{D1}\rightarrow -H_{D1}$ can be done by $\pi$ pulses on the qubits \cite{dl}, similar to the
spin echo method.

It is notable that $U(\gamma)$ operation proposed here possess the following distinct merits.
Firstly, it is insensitive to the thermal state of the cavity mode as the related influence
represented by the last two exponents in Eq. (\ref{U0}) is completely removed. Secondly, it
also remove the constrain of large detuning. It is noted that $T \sim 1/\eta$ for $\delta\sim
\eta$, which is comparable to the resonant coupling strategy and is much faster than that of
the conventional dispersive coupling with $\delta\gg \eta$. This fast gate operation is very
important in view of limited coherence times. Thirdly, it is obvious that the time needed for
this gate operation is comparable to a two-qubit gate as $A(t)$ is independent on $N$, and thus
the gate speed is not slowed down with the increasing of qubits. Therefore, it is readily
scalable to multiqubit scenario.

Now, we  illustrate how to generate the GHZ state in this system, which has many application in
quantum communication \cite{m}. By choosing $ \gamma =\pi/2$ and an initial state
$|\Psi\rangle_i=|00\cdots 0\rangle$ for $N$ NV centers, the final state
$|\Psi\rangle_f=\exp{\left(i\frac{\pi}{2}J_x^2\right)}|\Psi\rangle_i$ is found to be a GHZ
state given by \cite{ms}
\begin{equation}\label{GHZ}
|\Psi^f\rangle=\frac{1}{\sqrt{2}}[e^{-i\frac{\pi}{4}}|00\cdots
0\rangle+e^{i\pi(\frac{1}{4}+\frac{N}{2})}|11\cdots
1\rangle],
\end{equation}
when $N$ is even.   Note that $ \gamma =\pi/2$ can be achieved when choosing the parameter as
$\delta=2\sqrt{k} \eta$. Then, for $k=1$, one obtains $\delta=2\eta$ and the entanglement
generation time as $T=\pi/\eta$. For odd $N$, one can get GHZ state by applying another unitary
operator $U=\exp(-i{\pi \over 2} J_x)$ in addition to Eq. (\ref{ut}).

In the derivation of effective Hamiltonian in Eq.~(\ref{Hint1}), we have neglected the
following terms
\begin{eqnarray}\label{neg}
H_n &=&\frac{\eta}{2}  a  e^{-i\delta t} \sum _{j=1}^{N}
\left(e^{i\Omega  t}|+\rangle_j\langle-|-e^{-i\Omega t}|-\rangle_j\langle+|\right)+ \text{H.c.}   \notag \\
&=&\frac{i\eta}{2}  a  e^{-i\delta t} \sum _{j=1}^{N}
\left[\sin(\Omega t)\sigma^z_j-\cos(\Omega
t)\sigma^{y}_j\right]+\text{H.c.},
\end{eqnarray}
which could reduce the fidelity of the generated state. However, for $\sigma_z$ terms, one can
use the spin-echo technique to eliminate the infidelity \cite{dl}. Following Ref. \cite{ms}, we
consider the effect of the $\sigma_y$ terms on the gate operation. In the interaction picture
with respect to our free Hamiltonian $H_{free}=\delta a^+ a + \sum_{j=1} ^N {\Omega \over 2}
\sigma_j^x$, the interaction Hamiltonian is $H_{n,I}(t)=U^{\dagger}(t)H_{n} U(t)$ with
$U(t)=\exp{\left(-i H_{free} t\right)}$,  and we have the propagator $U'(t)$ from the Dyson
series. During the integration, one can treat $U(t)$ as a constant  as $H_n(t)$ is oscillating
much faster than the propagator. Then, under this assumption, we get
\begin{eqnarray}
U'(t) &=& 1-\frac{\eta}{2\Omega}\sin(\Omega t)\sum_{j=1}^{N}
U^{\dagger}(t)\sigma_j^yU(t)\\
&&+i\frac{\eta^2}{4\Omega^2}\sum\limits_{k\neq j}^{N} [1-\cos(2\Omega
t)]U^{\dagger}(t)\sigma_j^y\sigma_k^yU(t)\}  +\cdots.\notag
\end{eqnarray}
At the time $\delta t=2n\pi$,  the infidelity is
\begin{equation}
F_{in}=\xi[1-\cos(2\Omega t)],
\end{equation}
where $\xi= N(N-1)\eta^2/(8\Omega^2)$. Noted that the estimation of $F_{in}$ is obtained in the
interaction picture. Return to the rotating frame, the time evolution operator will get the
following additional term of ${U}_x=\exp(-i\sum _{j=1}^N {\Omega \over 2} \sigma_j^x t)$.
Therefore,  control the duration of classical microwave field accurately to fulfill  $\Omega
t=4n\pi$, the effect of the $\sigma_y$ terms and $U_x$ will both vanishes. Note that $\delta
t=2n\pi$, so $\Omega/\delta=2n$. Considering the further requirement of $\Omega \gg \delta$, we
may choose $\Omega = 6 \delta$.

To see the feasibility of the present scheme with current technology, we now use typical values
of physical parameters in the system to estimate  the operation time $T$.  NV centers are
located near the microcavity surface in order make the coupling between the NV center and WGM
maximum, which could be $G_{\max }\simeq 2\pi \times 1$ GHz with the transition wavelength
between the states $\left\vert e\right\rangle $ being 637 nm, the spontaneous decay rate of the
excited state being $\Gamma _0=2\pi \times 83$ MHz, and the mode volume be $V_{m}=$100 $\mu
m^{3}$ \cite{NJ}. For $\delta \ll \Delta$,  $\eta \simeq 2G \Omega _{L}/\Delta=2\pi \times 50$
MHz, with $\Omega_{L}=2\pi \times $ 500 MHz and $\Delta=2\pi \times $ 20 GHz. The effective
qubit spontaneous emission rate is estimated as \cite{Spo} $\Gamma _{eff}=\Gamma_{0}\Omega_L
G/\Delta ^{2}\approx 2\pi \times 0.1$ MHz. In the fused-silica microsphere cavity, the small
radius of 10 $\mu$m could lead to the $Q$ factor exceeding 10$^{9}$ with possible improvement
as demonstrated in \cite{ver}. The cavity decay rate rate is $\kappa =\omega _{c}/Q=2\pi \times
0.47$ MHz for $Q=10^{9}$. In our scheme, for $\delta = 2 \eta=2\pi \times 100$ MHz, the assumption
$\delta \ll \Omega$ with $\Omega=2\pi \times 600$ MHz \cite{omega} and $\delta \ll \Delta$ is
well-satisfied, then the entangled state generation time is $T =10$ ns, which is negligible
small compare to both the cavity decay and qubit coherence time  even at room temperature.

\begin{figure}[bp]
\centering
\includegraphics[width= 7cm]{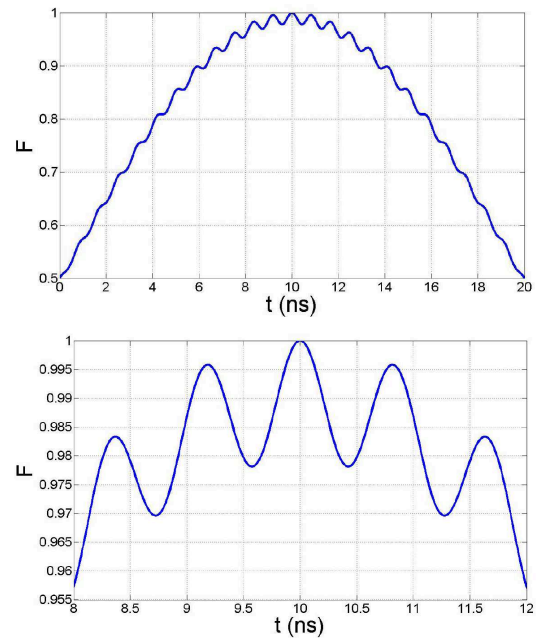}
\caption{Fidelity of the entanglement generation, see the main text for the chosen parameters.}
\end{figure}

With the above parameters, for $N=4$,  $\xi \approx 0.01$, and thus $F_{in} < 2.1 \%$,  which
increases with the increasing of \emph{N}. To see the time dependence of the fidelity of the
generated entangled state, as shown in Fig. (2), we numerically simulated the time-evolution of
the effective Hamiltonian of Eq. (11) plus $H_n$ in Eq. (14). We have defined the  fidelity as
$F=\langle \Psi^f|U(\gamma)|\Psi^f\rangle \times (1-F_{in})$. From the lower one of Fig. (2),
we can see that the generation is now very robust to small deviation of the  time duration.

In summary, we have proposed a scheme to  generate GHZ state of nitrogen-vacancy centers
coupled to a whispering-gallery mode cavity. The  fast scheme is  not sensitive to both the
spontaneous emission decay of the NV centers and thermal state of the cavity. Finally, the
scheme is readily scalable to multiqubit scenario as the generation time is independent on the
number of involved qubits.

\bigskip


This work was supported by the NFRPC (No. 2013CB921804),
the NSFC (No. 11004065),  the PCSIRT, and the NSF of Guangdong province (No. S2011040000403).

\end{document}